\documentclass[10pt,conference]{IEEEtran}
\usepackage{cite}

\ifCLASSINFOpdf
\else
\fi

\usepackage{graphicx}

\usepackage{url}

\begin{document}
%
\title{Engineering Formality and Software Risk in Debian Python Packages}

\author{\IEEEauthorblockN{Matthew Gaughan}
\IEEEauthorblockA{Northwestern University\\
Email: gaughan@u.northwestern.edu}
\and
\IEEEauthorblockN{Kaylea Champion}
\IEEEauthorblockA{University of Washington\\
Email: kaylea@uw.edu}
\and
\IEEEauthorblockN{Sohyeon Hwang}
\IEEEauthorblockA{Northwestern University\\
Email: sohyeonhwang@u.northwestern.edu}}


%


\maketitle

\begin{abstract}

While free/libre and open source software (FLOSS) is critical to global computing infrastructure, the maintenance of widely-adopted FLOSS packages is dependent on volunteer developers who select their own tasks. Risk of failure due to the misalignment of engineering supply and demand --- known as underproduction --- has led to code base decay and subsequent cybersecurity incidents such as the Heartbleed and Log4Shell vulnerabilities. FLOSS projects are self-organizing but can often expand into larger, more formal efforts. Although some prior work suggests that becoming a more formal organization decreases project risk, other work suggests that formalization may increase the likelihood of project abandonment. We evaluate the relationship between underproduction and formality, focusing on formal structure, developer responsibility, and work process management. We analyze 182 packages written in Python and made available via the Debian GNU/Linux distribution. We find that although more formal structures are associated with higher risk of underproduction, more elevated developer responsibility is associated with less underproduction, and the relationship between formal work process management and underproduction is not statistically significant. Our analysis suggests that a FLOSS organization's transformation into a more formal structure may face unintended consequences which must be carefully managed. 


\end{abstract}


%
\IEEEpeerreviewmaketitle

\section{Introduction}

Across global computing infrastructures, free/libre open source software (FLOSS) packages underpin the successful operation of widely used technical systems \cite{crowston_free/libre_2012}. Yet this crucial software is often ``underproduced'' --- that is, not as well-maintained as we might expect, given its importance \cite{champion_underproduction_2021}. Underproduction can be consequential, leading to 
disruption across global networks. Two notorious such failures are the FLOSS security defects known as Heartbleed and Log4Shell. The Heartbleed vulnerability impacted the widely-used OpenSSL library. At the time the vulnerability was announced, OpenSSL was decaying from community abandonment, with no full time developers and only \$2000 in annual donations. Heartbleed was eventually remediated by large-scale, formally organized developer engagement --- the very kind which OpenSSL had previously lacked \cite{walden_impact_2020}. The Log4j library is maintained by The Apache Software Foundation and a dedicated development team. In 2021, researchers identified the zero-day Log4Shell vulnerability, which enables attackers to inject malicious code into already-running Java programs \cite{noauthor_log4shell_2021}. While the Log4j team was able to respond quickly with a series of patches, the long-undetected nature of this vulnerability suggests that like OpenSSL, Log4j was underproduced.  

With 96\% of software in the `critical infrastructure sector' containing FLOSS code, the United States Cybersecurity and Infrastructure Security Agency (CISA) established an ``Open Source Software Security Roadmap'' in September 2023 \cite{noauthor_cisa_2023} highlighting the need to study the social and technical antecedents of FLOSS software failure. This national attention is timely: security research firm Sonatype found in their 2023 \textit{State of the Software Supply Chain Report} that across four major engineering ecosystems (NPM, Maven, PyPi, nugent) only 11\% of projects were actively maintained \cite{krill_report_2023}. Heightened scrutiny of FLOSS engineering practices as well as escalating warning signs of underproduction make examining the causes of underproduction all the more urgent.

A promising direction to diagnose both causes of and remediation strategies for underproduction is examining a structure commonly employed in FLOSS projects: commons based peer production (CBPP) \cite{benkler_wealth_2006}. Benkler defines CBPP as a production model reliant on decentralized individuals who self-select tasks, motivated by a range of factors \cite{benkler_coases_2002}. Yet Benkler also writes that in the face of ``knowledge intensive, creative, and complex'' problems, projects must organize to ``[elicit] diverse pro-social motivations'' from contributors \cite{benkler_peer_2017}. Said simply, although CBPP is an extremely valuable way of collaborating, FLOSS organizations may find it difficult to ameliorate underproduction because requisite yet undesirable tasks may lay neglected when contributors work according to their own interests. 

In this paper, we make three contributions: we offer empirical assessments of the relationship between (1) underproduction and overall governance formality, (2) underproduction and the concentration of developer responsibility, and (3) underproduction and the management of work products. We place our work in the context of previous work in §\ref{sec:background} and describe our analysis in §\ref{sec:methods} before presenting results of our analysis in §\ref{sec:results}. We discuss the implications of this work in §\ref{sec:discussion} with limitations noted in §\ref{sec:limitations} before concluding in §\ref{sec:conclusion}.

\section{Background}
\label{sec:background}

\subsection{Governance in CBPP}

Governance refers to the totality of the systems in an organization which incorporate decision making, operational control, and incentives \cite{yin_strategygovernance_2004}. Thus, as an organization matures, institutionalization in governance can be understood as ``the processes by which the social processes obligations, or actualities come to take a rulelike status in social thought or action'' across subunit actors (in the case of FLOSS project engineering, software developers) 
\cite{meyer_institutionalized_1977}. In CBPP governance, project institutionalization entrenches either formality or informality in decision-making processes \cite{healy_ecology_2003}.

The influential work of Ostrom on governing commons-based resources notes that communities prefer internally developed governance practices, which naturally leads to variations in the resulting governance arrangements \cite{ostrom_governing_2015}. For example, an analysis by Hwang and Shaw of CBPP governance on Wikipedia found that communities with shared goals, technical infrastructures, organizational structures, and institutional trajectories end up producing diverging rule sets and deliberating at length over shared rules \cite{hwang_rules_2022}. At times, how CBPP communities self-govern can lead to counter-intuitive patterns of institutionalization that introduce bureaucratization and hierarchies and create later barriers to participation, following Robert Michel's ``iron law of oligarchy.'' \cite{shaw_laboratories_2014}. The full extent of the productive repercussions of the variety of approaches to CBPP remains unclear.

\subsection{Governance of FLOSS Engineering}

Similar to other CBPP communities, FLOSS projects adhere to a wide range of organizational forms; the management of functional processes (leadership elections, funding allocation, onboarding, conflict resolution) is often internally defined with governance documents such as project constitutions and charters \cite{tourani_code_2017,de_laat_governance_2007}. Core to FLOSS governance is the allocation of developer labor and permissions (i.e., code integration, and how work products are released). For example, in centralized projects, formal project leadership may retain engineering management power, exemplified by Linus Torvald's sole control of code merging into the Linux kernel \cite{jiang_will_2013}.


Code merge patterns---who gets their work merged, and who decides---are a key indicators of how a project is being governed. Even in technically flat communities, Onoue et al. frame participant engagement in FLOSS projects as a hierarchy, with differences in who can enact development actions such as commits, pull requests, comments, and issue events  \cite{onoue_software_2014}. Thus, users who have the requisite permissions to merge code from peripheral forks into the primary development branch are empowered in project governance over those who simply commit to derivative branches. Tamburri et al. and van Meijel use the hierarchy of merge permissions as evidence of project hierarchy \cite{tamburri_organizational_2013,tamburri_exploring_2019,van_meijel_relations_2021}. De Stefano et al. adopt this approach to conclude that formality in internal engineering governance structures is associated with diminished community contribution to project code bases \cite{de_stefano_impacts_2022}. 

Tools like project progress trackers also give insight into FLOSS development. The management of such work products are often disputed within project communities; Crowston et al. observes that that there is no common patterns to how FLOSS projects employ public releases \cite{crowston_free/libre_2012}. 
GitHub ``milestones'' are indicators to mark the current state of a project, and are used by communities to track progress on collections of tasks such as issues or pull requests\footnote{https://docs.github.com/en/issues/using-labels-and-milestones-to-track-work/about-milestones}. While the milestone feature is platform-specific, GitHub remains the preeminent platform for FLOSS project hosting \cite{finley_for_2019}. Zhang et al. found a strong positive correlation between milestone usage and the count of other engineering measures, like commits and releases, in part due to the fact that older projects are more likely to use milestones than younger projects \cite{zhang_githubs_2020}.

\subsection{Governance and Underproduction}

Yin et al. observe that in canonical software engineering literature, the success of FLOSS software is primarily defined in two  ways: functional development processes or community cohesion \cite{yin_sustainability_2021}. Functional development processes include the management of project risk. In these terms, underproduction is a metric which considers potential for risk (developer neglect), its impact (user adoption), and the likelihood of risk resolution (code quality, bug resolution speed.) 
Champion and Hill identify underproduction by evaluating whether project quality is aligned with its importance when compared to a theoretical baseline of alignment, where importance and quality are aligned if their non-parametric rankings are the same \cite{champion_underproduction_2021}. For a set of packages from the Debian GNU/Linux distribution, Champion and Hill represent project importance through installation count and project quality as the average time to bug resolution (controlling for bug severity). 


Prior work examining FLOSS governance formalization vary in their measures of success and subsequent conclusions; Crowston and Howison even suggest that informality itself may be a metric of project success \cite{crowston_assessing_2006}. For critical FLOSS systems, Tamburri et al. defines success as ``24/7 availability in (a) fault-tolerant system'' \cite{tamburri_organizational_2013}. Yin et al. uses the advancement metrics of the Apache Software Foundation's incubator program to represent project success \cite{yin_open_2022}. De Stefano et al. focus on the frequency of product commits as a metric of community engagement and project success \cite{de_stefano_impacts_2022}. As such, Tamburri observes that as formality within software projects increases, projects may experience friction in the processes of developer engagement \cite{tamburri_organizational_2013}; Yin et al. conclude that isomorphic reproduction of formal governance structure may lead to project success \cite{yin_open_2022}; lastly, De Stefano et al. conclude that the more constrained a project's processes around merging are, the less engineering engagement it might receive \cite{de_stefano_impacts_2022}.

Observations that formalizing FLOSS governance may lead to both project abandonment and project success are not contradictory but warrant further empirical study. Overall, these findings suggest that projects with higher levels of formality are more likely to be underproduced. Due to the observations of De Stefano et al. \cite{de_stefano_impacts_2022} and Tamburri et al. \cite{tamburri_exploring_2019}, we propose: \textbf{$\mathbf{H_{A}}$: projects with higher formality are more likely to be underproduced}. Moreover, considering the findings from prior work on merge processes, we propose that 
\textbf{$\mathbf{H_{B}}$: projects with less concentrated developer responsibility are more likely to be underproduced}. Given the argument in Yin et al. that the governance of project work assists the development process \cite{yin_open_2022}, we propose \textbf{$\mathbf{H_{C}}$: projects with formal work process management are less likely to be underproduced.}

\section{Methods}
\label{sec:methods}

\subsection{Empirical Setting}
GNU/Linux distributions, includingDebian in particular, have a substantial history as settings for understanding software engineering communities \cite{spaeth_sampling_2007}, including their effective governance \cite{oneil_cyberchiefs_2009,sadowski_transition_2008} and lifecycles \cite{nguyen_life_2012}.
We examine a collection of projects packaged via the Debian GNU/Linux distribution, which is also a packaging source for Ubuntu and several other distributions. 
All projects were tagged by Debian developers as being implemented in the Python language, and all have repositories on GitHub. Therefore, these packages are part of one of the most widely-used GNU/Linux operating systems worldwide, with an upstream language and source code management platform in common.

\subsection{Data}

Our unit of analysis is the software project, n = 182. All projects meet three criteria: they are in the Debian GNU/Linux distribution, included in the dataset published by Champion and Hill\cite{champion_underproduction_2021}, and tagged by Debian maintainers as being written in Python. Due to the manual identification of upstream repositories, our data set was constrained to Debian packages written in the Python language; further research is necessary to study the Debian package ecosystem writ large. The median project in this dataset was 13 years old, with 44 members in their developer communities. 

To identify the upstream repository of each package that was the same as the one used by Debian, we first examined Debian metadata: in the Ultimate Debian Database \cite{nussbaum_ultimate_2010}, on the Debian package website\footnote{\url{https://packages.debian.org/}}, and inside the package as stored in Debian's GitLab instance\footnote{\url{https://salsa.debian.org/}}. If no upstream repository location was identified, we examined the files associated with the package (README.txt if available, and the documentation and license files if not).We then collected package commit and milestone history using the CHAOSS GrimoireLab Perceval tool \cite{duenas_perceval_2018} and the public GitHub API, respectively. We bounded our data collection within the range of 2/8/2008 - 11/09/2023; February 8, 2008 was the date of GitHub's founding.\footnote{Our data and code are available on the Harvard Dataverse: https://doi.org/10.7910/DVN/WENTBH}


\subsection{Measures}

We operationalize project \textit{governance formality} ($H_{A}$) using Tamburri et al.'s YOSHI formality score \cite{tamburri_organizational_2013}, one of six metrics addressing project governance (alongside community structure, geodispersion, longevity, engagement, and cohesion). YOSHI has been used extensively in prior empirical studies of FLOSS governance \cite{tamburri_exploring_2019,catolino_secret_2019, van_meijel_relations_2021, mauerer_search_2022, de_stefano_impacts_2022}.

\begin{equation}
\label{formality-eq-orig}
    \textrm{Formality Score} = \frac{MMT}{MS / LS}
\end{equation}

This measure considers developer responsibility, work processes, and age as component variables to calculate overall project formality, shown in Equation \ref{formality-eq-orig}. \textit{Mean Membership Type} (MMT) represents developer responsibility. This is further explained below with Equation \ref{original-mmt-equation}. MS refers to work processes, in our case GitHub milestone count, while LS refers to project lifespan in days. The potential range of scores is between 0 and 10,000. However, because this definition would filter out projects who do not use milestones entirely---a substantial portion of our sample--- we develop an augmented calculation of the formality score where the MMT is divided by whether or not the project employs milestones (represented by a binary 1:2 classification) (MSE) per the project's age grouping (AG), which is included as a control variable given age may generally impact formalization. Projects were grouped in bins of (1) 0-9 years old (n=44), (2) 9-12 years old (n=44), (3) 12-15 years old (n=49), and (4) 15-16 years old (n=64). The breaks were selected to create bins of similar sizes. Each project was coded with a corresponding group label of one to four. This augmented formality measure is shown in Equation \ref{formality-eq} and the resulting metric ranges between 0 and 10.

\begin{equation}
\label{formality-eq}
    \textrm{Augmented Formality Score} = \frac{MMT}{MSE/ AG}
\end{equation}

To evaluate \textit{concentration of developer responsibility} ($H_{B}$), we draw on a measure used in the overall formality score mentioned above: MMT, which represents the share of project developers who hold more responsibilities via their privileged roles in the engineering process, such as merge permissions. As seen in Equation \ref{original-mmt-equation}, this metric is a weighted average of collaborators in a given FLOSS community, with the community defined as all individuals whose committed edits to project files have been accepted in the main branch \cite{tamburri_organizational_2013,tamburri_exploring_2019}. Collaborators are developers who have authored a merge into the primary code branch and thus have the requisite permissions to do so, contributors are commit authors who have never merged into the primary code branch \cite{van_meijel_relations_2021}. Higher MMT averages may suggest flatter organizational structure, since it indicates widely diffused merge activities \cite{gousios_exploratory_2014}.

\begin{equation}
\label{original-mmt-equation}
   MMT = \frac{1}{|M|}\sum_{m \in M}^{}\left\{ \begin{array}{cl}
       2 & \textrm{If \textit{m} is a collaborator.} \\
       1 & \textrm{If \textit{m} is a contributor.}
       \end{array} \right.
\end{equation}


We measure \textit{formal work process management} ($H_{C}$) via projects' use of GitHub milestones \cite{tamburri_organizational_2013}, again drawing on the formality score. However, some projects in our dataset are hosted on platforms which lack the milestone metric or do not use it entirely, with prior work indicating only around 20\% of projects on GitHub used the milestone tool \cite{zhang_githubs_2020}. Thus, we employed two metrics of milestone usage. One was the numeric count of milestones that a given project was using; the other was a binary classification of whether or not a project used milestones. Around 25\% of packages studied in this data set used milestones.
To measure the \textit{underproduction factor} of a FLOSS project, we use the mean underproduction factor estimates published in Champion and Hill \cite{champion_underproduction_2021}. As previously described in \ref{sec:background}, the underproduction metric measures a project's engineering activity against the project's importance.






\subsection{Analytic Plan}

Linear regression is an established approach of identifying associative relationships between two continuous variables. In our evaluation of the continuous measures of $\mathbf{H_{A}}$ (formality score), $\mathbf{H_{B}}$ (responsibility concentration), $\mathbf{H_{C}}$(work processes), and project age, we used linear regression models to evaluate the relationship with the project's underproduction factor and evaluate significance at the $p < .05$ level. 

Given our constrained data sample, we also employed a statistical power analysis to evaluate the impact of data sample size on our results when relevant ($H_{A}$, $H_{C}$). 

\section{Results}
\label{sec:results}

\begin{table}[h]
\caption{This table displays the relationships between underproduction and three project metrics: formality, MMT, and milestones. Models for $H_{B}$ and $H_{C}$ include a control for age with factor variables; 0 to 9 years old is the baseline category.}
\centering
\setlength{\tabcolsep}{0.3\tabcolsep}
\begin{tabular}{l c c c }
\hline
 & ($H_{A}$) formality & ($H_{B}$) MMT & ($H_{C}$) milestones \\
\hline
(Intercept)         & $-0.78^{*}$       & $1.65^{*}$        & $-0.89^{*}$       \\
                    & $ [-1.27; -0.29]$ & $ [ 0.06;  3.25]$ & $ [-1.32; -0.45]$ \\
Augmented formality & $0.17^{*}$        &                   &                   \\
                    & $ [ 0.05;  0.29]$ &                   &                   \\
MMT                 &                   & $-1.38^{*}$       &                   \\
                    &                   & $ [-2.21; -0.54]$ &                   \\
Age 9-12y &                   & $0.07$            & $0.27$            \\
                    &                   & $ [-0.53;  0.67]$ & $ [-0.33;  0.88]$ \\
Age 12-15y               &                   & $0.60^{*}$        & $0.79^{*}$        \\
                    &                   & $ [ 0.01;  1.18]$ & $ [ 0.20;  1.38]$ \\
Age 15-16y               &                   & $1.15^{*}$        & $1.53^{*}$        \\
                    &                   & $ [ 0.55;  1.74]$ & $ [ 0.96;  2.11]$ \\
Milestone count     &                   &                   & $0.01$            \\
                    &                   &                   & $ [-0.13;  0.15]$ \\
\hline
R$^2$               & $0.04$            & $0.22$            & $0.17$            \\
Adj. R$^2$          & $0.04$            & $0.20$            & $0.15$            \\
Num. obs.           & $182$             & $182$             & $182$             \\
\hline
\multicolumn{3}{l}{\scriptsize{$^*$ Null hypothesis value outside the confidence interval.}}
\end{tabular}
\label{tab:underprod-rel}
\end{table}



\subsection{$H_{A}$: Formality Score}


We found a statistically significant relationship between a project's score and underproduction factor ($p < .005$). However, the relationship between formality score and underproduction was relatively small, where a unit increase in formality score was associated with only a 0.17 increase in underproduction factor. 
This result is evidence in favor of our hypothesis $H_{A}$, higher formality is associated with increased risk of underproduction.

The analysis reported in Table \ref{tab:underprod-rel} uses the augmented formality calculation in Equation \ref{formality-eq}, which enabled us to include projects who did not use GitHub milestones, giving us a larger sample size (n=182). 

Using the original formulation of the formality score in Equation \ref{formality-eq-orig} would have limited our analysis to projects whose milestone counts are greater than zero; the sample size for this model was 44. Testing $H_{A}$ using this alternate metric found no statistically significant relationship between the score and a project's underproduction factor. Given our small sample, we conducted a power analysis to assess the impact of sample size in detecting an effect size of at least \(\beta\) = 0.00017 (this is the significant effect size of the augmented formality score, but scaled commensurately with the original score's range.) 
To simulate our data, we assumed that the formality score encapsulated other metrics such as MMT, milestones, and age and that formality score data follow a beta(\(\alpha\):1, \(\beta\):3) distribution. 
After running 1000 simulations on data sets of size 75, we are able to reject the null hypothesis. 
The power analysis provides evidence that, to the extent that this simulated pilot data is representative of the population, our sample size is insufficient to conclude the original formality score's relationship to underproduction, and lends validity to our decision to use the augmented formality calculation displayed in Equation \ref{formality-eq}. 

\subsection{$H_{B}$: Concentration of Developer Responsibility}

\begin{center}
    \begin{figure}[h!]
        \includegraphics[width=\linewidth]{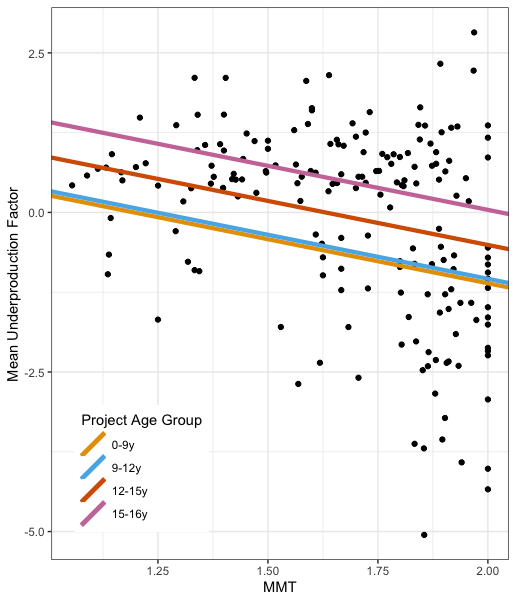}
        \caption{Plot showing the relationship between projects' MMT and mean underproduction factor. Plotted lines are drawn from the results of our model for $H_{B}$.} \label{fig:mmt-underprod}
    \end{figure}
\end{center}

A linear regression model fit with MMT values indicated strong statistical significance for a negative relationship between MMT and mean underproduction values, suggesting that as the share of project developers who assume privileged roles (MMT) increases, the factor of underproduction decreases. The results are statistically significant ($p < .002$), where a unit increase in MMT is associated with a 1.38 point decrease in underproduction risk. This result is contrary to our hypothesis $H_{B}$, and instead provides evidence that increasing dispersion of responsibility is associated with lower underproduction risk, rather than higher. The effect size (-1.38) is relatively large, given that underproduction scores for our data only range between -5.05 and 2.81, suggesting that this association is also practically significant.
Figure \ref{fig:mmt-underprod} displays the relationship between projects' MMT and mean underproduction factor. 

\subsection{$H_{C}$: Formal Work Process Management}


A linear regression model fit with projects' mean underproduction scores and project usage of GitHub milestones did not point to a statistically significant relationship between the two measures ($p = 0.09$). However, as with formality score, a lack of significance may be due to our relatively small sample size (n = 182). Thus we conducted a power analysis simulation to assess whether we had sufficient observations to detect an effect size of at least \(\beta\) = 0.40; a ~5\% impact in underproduction factor within our data set. To simulate our data, we transposed MMT into a 0-1 range, modeled the measure's established relationship of -1.38 units of underproduction factor, and assumed MMT data follow a beta (\(\alpha\):5, \(\beta\):1) distribution. From initial analysis of collected data, we also assumed that milestone data follow a binomial distribution with probability 0.247. Running 1000 simulations of data sets of n=300 we were able to reject the null hypothesis. This provides evidence that, to the extent that our pilot data and simulation are representative of the population, our sample size is insufficient to conclude there is no meaningful relationship between formal work process management (GitHub milestones) and underproduction.

\section{Discussion}
\label{sec:discussion}

Overall, our results paint a nuanced picture of the relationship between formalization and underproduction in FLOSS projects. Our analysis of formal structure ($H_{A}$) using the augmented formality score showed that overall, more formal structures were associated with a minor increase in underproduction risk. These results support arguments from prior work from Tamburri et al. \cite{tamburri_organizational_2013} and De Stefano et al. \cite{de_stefano_impacts_2022} that formal governance concentration is related to increased project risk.

However, our analysis of the distribution of privileged roles across developers ($H_{B}$, using MMT) showed that an increase in project developers with privileged roles is associated with decreased underproduction risk. Examining the relationship in Figure \ref{fig:mmt-underprod} suggests that this effect occurs when MMT is especially high, given the relatively flat relationship when MMT ranges between 0-1.75. One reason higher MMT may be associated with lower underproduction risk may be that the diffusion of privileged roles across more developers may indicate higher commitment to the project overall. For example, cursory analysis of project age shows a small positive relationship between age and a project's risk of underproduction. As a package ages, not only is there a higher likelihood of software adoption, there is also a higher likelihood of community abandonment. While we cannot make any causal claims, our results suggest the diffusion of project responsibility may help reduce risk of community abandonment and thus, risk of underproduction.

Finally, our analysis of formal work processes as captured by project milestone usage did not yield statistically significant results. However, our power analysis suggests that this was due to sample size, and we believe this is a valuable direction future work to better understand the effects of formalization.

\section{Limitations and Future Work}
\label{sec:limitations}

In this project, we only examine Python projects which are within the Debian distribution and for which an underproduction factor measure was available. Our analysis was further limited to projects which are hosted on platforms supporting our data collection approach (GitHub/GitLab); while these platforms are widely used hosting platforms for FLOSS projects, these limitations restricted our sample size (n=182). Although our work offers insight into risk around the widely-used Debian distribution, expanding this sample is an important direction of future work. Moreover, the metrics we employ do not account for longitudinal changes in project governance; this remains a topic of further research.



\section{Conclusion}
\label{sec:conclusion}
FLOSS organizations face numerous challenges as they seek to mobilize volunteers to produce secure, high-quality software. We examined three hypotheses about the impact of formality on underproduction risk, finding that project formality may be a relevant indicator of higher software risk, that the diffusion of engineering responsibility was associated with lower risk, and that work product management is an uncertain predictor. Taken together, these results suggest governance informality and broader sharing of responsibility are beneficial to FLOSS projects. Given the importance of FLOSS across software ecosystems, unlocking the complex relationship between FLOSS governance and underproduction offers a new avenue for addressing the cybersecurity risks facing digital infrastructure.  


\section*{Acknowledgment}
This work is indebted to the volunteer developers producing FLOSS who have made their work available for inspection. We also gratefully acknowledge support from the Sloan Foundation through the Ford/Sloan Digital Infrastructure Initiative (Sloan Award 2018-113560 and the National Science Foundation (Grant IIS-2045055). This work was conducted using research computing resources at Northwestern University.



\bibliographystyle{IEEEtran}
\bibliography{references.bib}
\end{document}